\documentclass[sigconf]{acmart}

\usepackage{booktabs} 
\usepackage{graphicx}
\usepackage{caption}
\acmDOI{}

\acmISBN{}

\acmConference[]{WWW `19 FATES}{San Francisco, CA}

\begin{document}
\title{Trust and Trustworthiness in Social Recommender Systems}

\author{Taha Hassan}
\affiliation{%
  \institution{Computer Science Department, Virginia Tech}
  \city{Blacksburg}
  \state{VA}
}
\email{taha@vt.edu}

\author{D. Scott McCrickard}
\affiliation{%
  \institution{Computer Science Department, Virginia Tech}
  \city{Blacksburg}
  \state{VA}
}
\email{mccricks@vt.edu}

\renewcommand{\shortauthors}{T. Hassan}

\begin{abstract}
The prevalence of misinformation on online social media has tangible empirical connections to increasing political polarization and partisan antipathy in the United States. Ranking algorithms for social recommendation often encode broad assumptions about network structure (like homophily) and group cognition (like, social action is largely imitative). Assumptions like these can be na{\"i}ve and exclusionary in the era of fake news and ideological uniformity towards the political poles. We examine these assumptions with aid from the user-centric framework of \textit{trustworthiness} in social recommendation. The constituent dimensions of trustworthiness (diversity, transparency, explainability, disruption) highlight new opportunities for discouraging dogmatization and building decision-aware, transparent news recommender systems. 

\end{abstract}


\maketitle
\section{Introduction}

\begin{figure}
\includegraphics[width=0.45\textwidth]{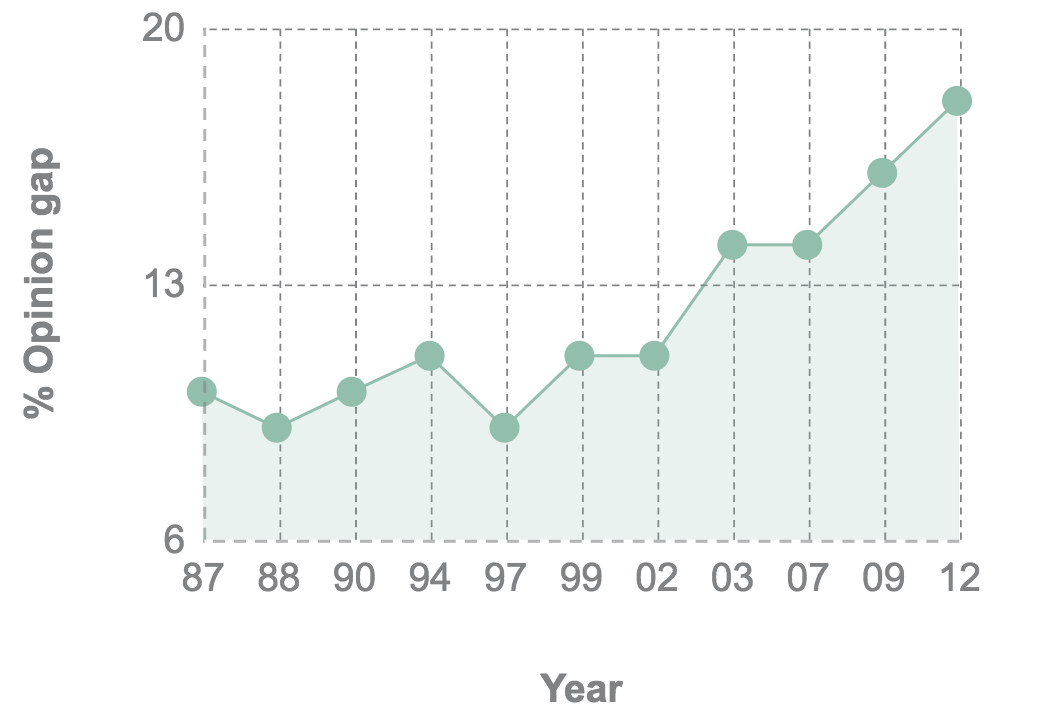}
\caption{Partisan differences in political values (average percentage difference between Republicans and Democrats) on a 40-point values questionnaire \cite{pewmediahabits}}
\label{fig:gap}
\vspace{5pt}
\includegraphics[width=0.48\textwidth]{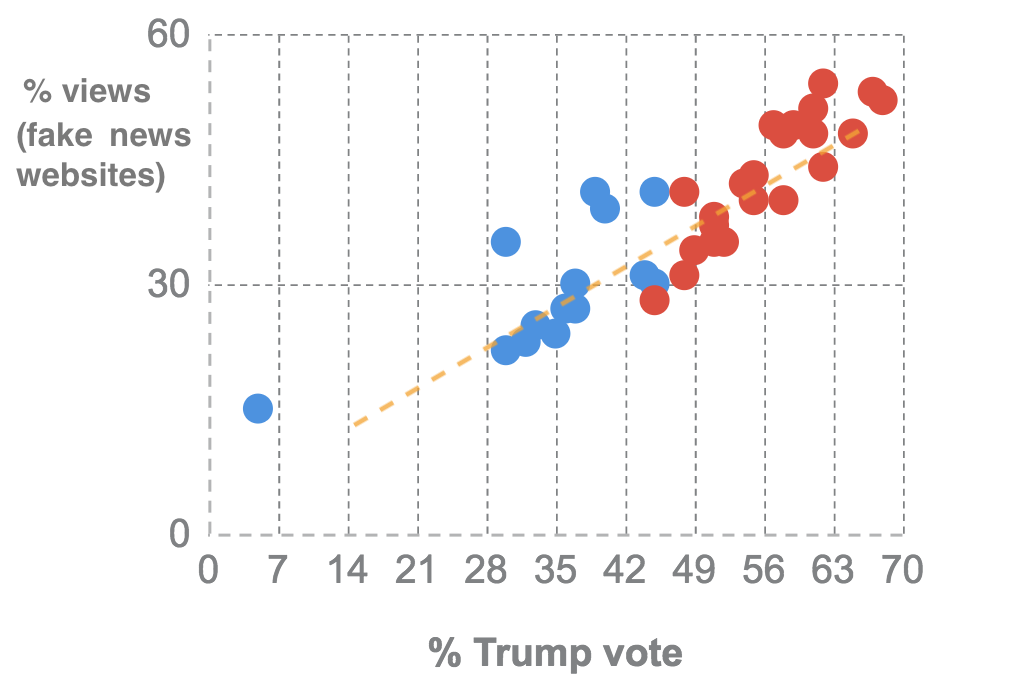}
\caption{Voting behavior (by state) and average daily viewership of visitors to websites serving fake news. Blue represents a state-wide Democratic victory, red represents a Republican victory \cite{fourney}}
\label{fig:conn}
\end{figure}

In the 2016 US Presidential Election, social media was one of the key venues for dissemination of fake news, with aggregate viewership of fake news strongly correlated with aggregate voting patterns \cite{fourney}. These patterns hint at deeper geographical and sociopolitical groupings in the US (see figure \ref{fig:gap}). Differences in media habits by political affiliation have also been the subject of attention of late. A Pew Research Center study \cite{pewmediahabits} found little evidence of overlap between news sources frequented and trusted by self-identified consistent liberals and conservatives. Conservatives gravitated towards a small number of news outlets, often just one, and exhibited distrust of a majority of the mainstream news sources. Liberals consulted a wider variety of news outlets, with social issue-based choices. In addition, consistent liberals were more likely to `unfriend' or block someone because of their political views, while consistent conservatives were more likely to have friends who agree with their political views. 
According to the Pew Center study, Facebook draws twice as many consumers of news on politics and government relative to Yahoo or Google News, and about 40\% of the US population would seek these news stories on Facebook week before the survey. A number of social recommender systems serving content as varied as pages, groups, events and jobs, work behind-the-scenes at Facebook. User interaction with a social recommender encodes interesting technical, behavioral and demographic trends, especially in the era of increasing partisan rifts. This study explores key limitations of social recommendation from a critical-theoretic standpoint and provides new design considerations for their future.

The next sections review the conventional notion of trust-awareness, present a critique of this notion within the context of political polarization and explore issues of transparency, explainability and diversity for social news recommenders.

\section{Related Work}
Implementations of recommender systems have been traditionally attributed to one of two approaches: the model-driven and the data-driven \cite{koren}. The first, known as collaborative filtering (CF) refers to modeling the users and items jointly, by a shared space of latent feature vectors \cite{hasancollaborative}. The second, neighborhood modeling (NM), identifies user-user and item-item similarities directly. In practice, these two approaches have similar ancestry and can be put together in hybrid approaches. 

Trust-based recommender systems are concerned with learning the preferences of trustworthy social neighbors, or 'friends' of an individual user, as well as the mistrusted 'foes' \cite{learningtorank}. These preferences inform the latent features inferred in CF, such that features for an individual are ranked closer to his or her friends' features, rather than the foes. In \cite{massa}, learning-to-rank models minimize a loss function on a personalized ranking function. Using trust and mistrust relationships proves effective in combating the sparsity inherent in users' preferences data.

User-perceived quality of recommender systems' output is a broad way to evaluate anything from aggregate emotional impact to perceived relevance and variety. User-perceived variety or diversity in the recommender output is a related, important notion. Authors in \cite{orginterface} explore an organization interface (ORG) as opposed to a list interface for a top-N style recommender. This interface clusters and annotates subgroups of the recommender output. These annotations go beyond category labels and express tradeoffs in product quality and price. ORG ranked better than lists in perceived ease of use and diversity. Studies like \cite{diversification}, however, note that statistical accuracy of such recommendations might be lower, even as they are rated better in quality by field trial participants.

\section{Trustworthiness}
\subsection{A Critique of Trust} 
User feedback in recommenders, be it explicit (ratings, reviews) or implicit (click behavior, browsing habits) is often sparse and non-generalizable. This sparsity has driven the need for the conventional notion of trust. Often leveraged in CF, this notion of trust assumes homophily in the social network, that like connects with or favors like \cite{fourney}. Trust-aware CF, therefore, ranks the judgment of trustworthy friends as consistently higher than those of untrustworthy peers and foes. From a broad critical-theoretic standpoint, this construction of trust can be arbitrary and cynical. Note that sparsity encodes an emergent market need for showing relevant news stories rather than in-depth or disruptive ones. Also note that, the qualifier `emergent' helps avoid the trendy critique of reverse-cynicism. It refers to the aggregate of choices by a large number of people potentially exercising their free will, choices such as reviewing or rating items less frequently than idly browsing a news feed.

\begin{figure}
\includegraphics[height=1.6in, width=3.3in]{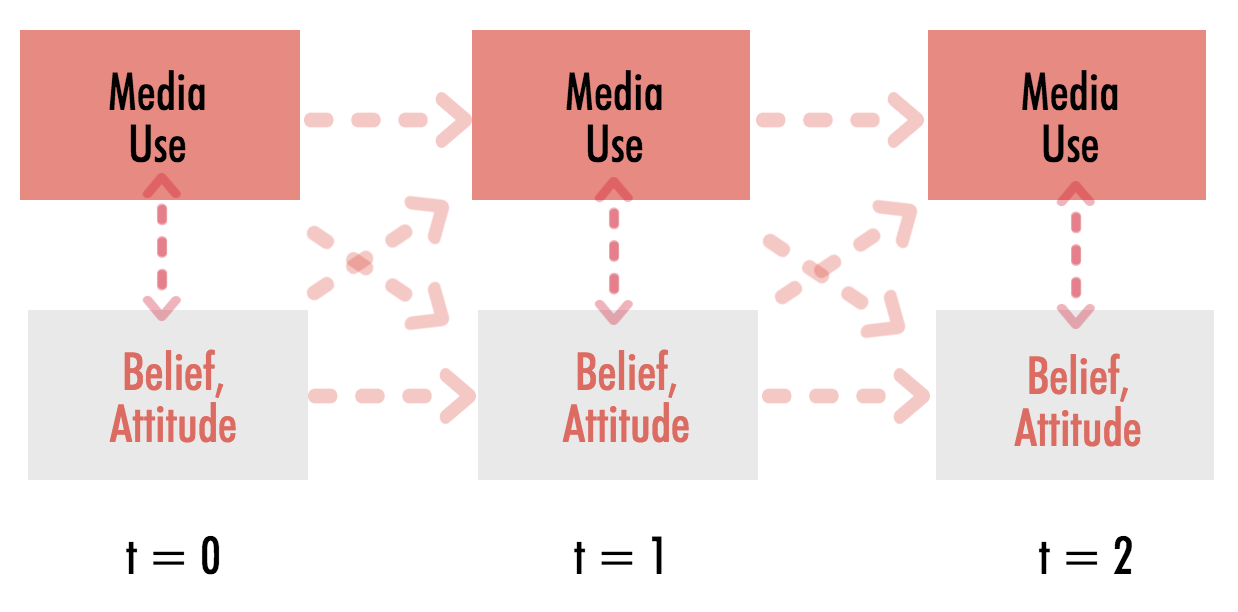}
\caption{A three-way, lagged model of behavior reinforcement spirals \cite{spirals}.}
\label{fig:spiral}
\end{figure}

To expand this critique, we can turn to a large volume of literature in the behavioral sciences domain \cite{influence}\cite{spirals} that examines the degree and scope of recommendation influence on online choices. Recall how consistent conservatives and liberals have strong selective tendencies in choosing whom to be friends with and what media sources to trust. The author in \cite{spirals} presents a theoretical framework for bias reinforcement in social media (see figure \ref{fig:spiral}). A closed feedback loop aims to maximize the use of a single social medium in the absence of competing interests. While a statistical analysis of these models is beyond the scope of the study, there are important conceptual aspects to consider here. One, consider the multitude of paths that can lead to feedback loops of gratification. Two, consider the assumptions behind nominal categories like media use, such as that of a closed system. In practice, institutions like marriage, family and religion, as well as competing media sources might help limit the scope of this reinforcement.

\begin{figure*}
\includegraphics[width=0.64\textwidth]{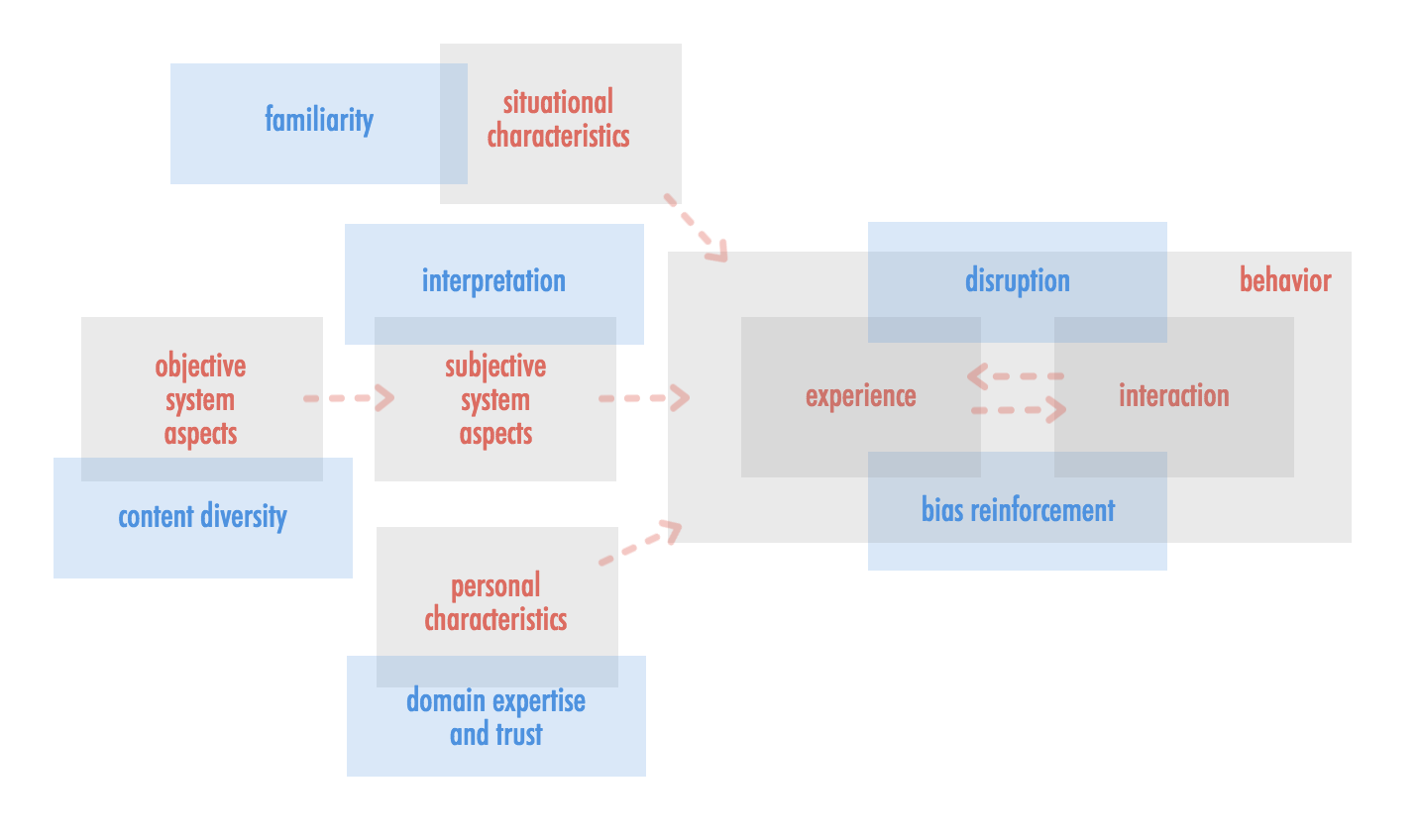}
\caption{A recommender systems UX framework \cite{knijnenburg2012} from the perspective of trust and disruption.}
\label{fig:framework}
\end{figure*}

While there is some prior work on longitudinal trends in recommender output and rating diversity \cite{filterbubble}, reinforcement loops have not been formally studied from a recommender UX perspective (to the best of the author's knowledge). It is helpful, therefore, to reflect on the interactions that facilitate such loops. Authors in \cite{knijnenburg2012} describe a user-centric framework for perceived effort, accuracy and satisfaction of a recommender systems user (see figure \ref{fig:framework}). A certain dual causation exists between \textit{experience} and \textit{interaction} in this UX model. \textit{Experience} can inform and modify \textit{interaction}, as much as \textit{interaction} informs and modifies \textit{experience}. We can thus, leverage the spiral model in \cite{spirals} to conceptualize bias-reinforcement as an undesirable byproduct of this duality. Furthermore, the study reports that the users who rate the recommendations as `effective' have on average, a lower total viewing time and total number of watched clips. They do not necessarily consume a lesser volume of information, but less of their time is spent browsing than it is on watching recommended clips. 

Another important distinction is the so-called `greedy' approach to search. Some users are looking to find the `best' possible recommendation while browsing, while others are looking to find the first perceived optimal recommendation. Domain expertise and trust affect these decisions \cite{knijnenburg2011}. In that, the tendency to favor sophisticated control (like, user-defined weights and order) over sort-only control correlates well with domain expertise. An `average' user is likely a lot more passive in taking a recommender system's output at face value. Authors in \cite{filterbubble} report that for long-term users of the MovieLens recommender, the diversity of recommender output and of explicit user ratings narrows over time. However, the drop in diversity is greater for users who tend to ignore the recommender output whilst rating content relative to those who do not.

This calls for design considerations that go beyond the local-neighborhood approach to recommendations. The output of a recommender system should help disrupt rather than reinforce feedback loops \cite{bz}. This involves a move towards explainability and transparency, as well as diversification and provocation. 

\subsection{Conceptual Dimensions of Trustworthiness}
In the broadest sense, a trustworthy recommender is a recommender that enables transparent and interpretable interaction against the rising current of dogmatization and partisan antipathy. The key conceptual dimensions are as follows.

\subsubsection{Diversification}
Diversification for a news recommender improves the potential for a given user's encounter with alternative perspectives on a given social issue and disrupts the feedback loops explored in the previous section. There is room for creating a notion of diversity using a quality ranking for each news article. Google News, for instance, features `in-depth' articles for a subset of search queries based on automated strategies. Similarly, Amazon.com recommenders arrange reviews by usefulness based in part on explicit feedback by readers of said reviews. A diversified recommender should recognize the degree of substantiation, i.e., it should be able to distinguish between an `in-depth' analysis or news story versus a `broad overview' and present a mix of labeled samples from each. 
Empirical studies in \cite{knijnenburg2011} and \cite{knijnenburg2012} associate the tendency to tag with a user's trust of the recommendation system and lack of privacy concerns. Therefore, manual tagging of article quality needs to be evaluated for agreement between independent labelers and between labelers and third-party fact-checking agencies.

\subsubsection{Disruption}
Model-level disruption is a more intrusive strategy that can potentially challenge the neighborhood assumption of recommender systems and create more provocative collections of news and analyses. The present state of recommender systems makes it difficult to get recommendations on a specified social perspective. In other words, we cannot ask a recommender system to fetch say an anarcho-syndicalist or a second-wave feminist perspective on a social issue or a news story, unless there is explicit or implicit history of a user browsing articles similar to such perspectives. However, consider the notion of transfer learning. Transfer learning for CF \cite{transferlearning} allows cross-domain application of machine-learnt models and sparsity reduction. There is, however, no transfer-learning framework that uses a user-specified target domain to rank and retrieve recommendations, to the best of the author's knowledge. 

\subsubsection{Explainability and Interpretation}
Another facet of explainability (as well as transparency) is decision-awareness for the user. Netflix's movie recommender, for instance, includes annotations on why it ranked a given movie or a TV show highly among the recommended content. These include references to actors, directors, genres and relevant decades, among others. A news recommender, especially one concerned with popular conversations, can potentially increase the users' trust and confidence in its output using explanatory annotations. This might especially be true with automated flagging of ad spam, content involving bullying and online predatory behavior, or NSFW images and video data. 

\section{Future Work}
This study explores the theoretical foundations of trust-aware ranking in social recommenders.  A key frontier in building such systems is verifying the user-perceived quality of disruptive interaction. Therefore, crowd-sourced tagging of article quality and authenticity, third-party verification, progressive disclosure of perspectives in a conversation and disruptive personas are subject of future empirical investigations by the author. 

\section{Conclusion}
Disruption in recommender systems can take one of many forms. However, the relatively new interest in this domain implies a lack of frameworks, empirical studies, field trials and ethnographic work. Culture-awareness is an ambitious frontier for recommender systems of the future, and one that encompasses ideas like model differentiation, language preservation and community building. Diversification, transparency and disruption are key building blocks for tools that enable these ideas.

\bibliographystyle{ACM-Reference-Format}
\bibliography{sample-bibliography}

\end{document}